\documentclass[showpacs,apl,reprint,amsmath,amssymb,superscriptaddress,nobibnotes,twocolumn, preprintnumbers, bibnotes]{revtex4-1}
\usepackage{graphicx}
\usepackage{bm}
\usepackage{epsfig}
\usepackage{amsmath}
\usepackage{color}
\usepackage{array}

\newcommand{\etal}{{\it et al.}}
\newcommand{\LiOsO}{LiOsO$_{3}$}

\begin{document}


\title{Pressure-induced enhancement of non-polar to polar transition temperature in  metallic LiOsO$_3$}

\author{Esteban~I.~Paredes Aulestia}
\author{Yiu~Wing~Cheung}

\affiliation{Department of Physics, The Chinese University of Hong Kong, Shatin, New Territories, Hong Kong, China}

\author{Yue-Wen Fang}
\email{Present address: Department of Materials Science and Engineering, Kyoto University, Kyoto 606-8501, Japan}
\affiliation{NYU-ECNU Institute of Physics, NYU Shanghai, Shanghai, 200062, China}


\author{Jianfeng He}
\author{Kazunari Yamaura }
\affiliation{Research Center for Functional Materials, National Institute for Materials Science (NIMS), 1-1 Namiki, Tsukuba, Ibaraki 305-0044, Japan}
\affiliation{Graduate School of Chemical Sciences and Engineering, Hokkaido University, North 10 West 8, Kita-ku, Sapporo, Hokkaido 060-0810, Japan}

\author{Kwing~To~Lai}
\affiliation{Department of Physics, The Chinese University of Hong Kong, Shatin, New Territories, Hong Kong, China}

\author{Swee~K.~Goh}
\email{skgoh@phy.cuhk.edu.hk}
\affiliation{Department of Physics, The Chinese University of Hong Kong, Shatin, New Territories, Hong Kong, China}
\affiliation{Shenzhen Research Institute, The Chinese University of Hong Kong, Shatin, New Territories, Hong Kong, China}

\author{Hanghui Chen}
\email{hanghui.chen@nyu.edu}
\affiliation{NYU-ECNU Institute of Physics, NYU Shanghai, Shanghai, 200062, China}
\affiliation{Department of Physics, New York University, New York  10003, USA}

\date{\today}


\begin{abstract}
LiOsO$_3$ undergoes a continuous transition from a centrosymmetric $R\bar{3}c$ structure to a polar $R3c$ structure at
$T_s=140$~K. By combining transport measurements and first-principles
calculations, we find that $T_s$ is enhanced by applied pressure, and it
reaches a value of $\sim$250~K at $\sim$6.5~GPa. The enhancement is
due to the fact that the polar $R3c$ structure of LiOsO$_3$ has a
smaller volume than the centrosymmetric $R\bar{3}c$ structure.
Pressure generically favors the structure with the smallest volume, and therefore
further stabilizes the polar $R3c$ structure over the $R\bar{3}c$
structure, leading to the increase in $T_s$.    
\end{abstract}


\maketitle


Ferroelectric materials are actively explored in both fundamental
science and applied research for their potential industrial
applications such as non-volatile memory devices, sensors and solar
cells
\cite{LiFei2018NatMater,Garcia2014nc,fang2017lattice,hsieh2016permanent,QiPRL2018,lu2018apl,LiuPRB2018,nakamura2017nc,
  Chen2017}. In ferroelectric materials, a continuous displacive phase
transition from a centrosymmetric structure to a non-centrosymmetric
structure occurs at Curie temperature $T_C$, below which a spontaneous
polarization develops.  Since itinerant electrons can effectively
screen internal electric fields, it is a conventional wisdom that
ferroelectric materials are insulators~\cite{Shi2013}. However,
Anderson and Blount \cite{Anderson1965} argued that a second-order
structural transition involving the formation of a polar axis and the
disappearance of the inversion center, analogous to that of displacive
ferroelectrics, may result in a `ferroelectric-like metal'. Puggioni
\etal\ \cite{Puggioni2014NC} further deduced that such transition may
be a result of the electrons at the Fermi level that are only weakly
coupled to the soft phonon mode that drives the second-order phase
transition into the polar structure.  Recently, Shi~\textit{et
  al.}~\cite{Shi2013} showed clear evidence of a polar metallic state
in \LiOsO, which has stimulated many further experimental and
theoretical
works~\cite{LoVecchio2016,kim2016Nature,Varjas2016PRL,Puggioni2014NC,
  xiang2015prb,
  HMLiuPRB2015,Puggioni2015PRL,Filippetti2016NatComm,Giovannetti2014,Yu2018}.  At a
critical temperature $T_s=140~$K, \LiOsO\ undergoes a structural
transition from centrosymmetric $R\bar{3}c$ (space group 167) at high
temperature to a polar, non-centrosymmetric $R3c$ (space group 161)
below $T_s$. This structural transition is analogous to that in
LiNbO$_3$, which is a well-known ferroelectric compound
($T_C=1483$~K~\cite{Nassau1965}) and thus the insulating equivalent
of \LiOsO.  The heat capacity data displays a broad peak at $T_s$
without any thermal hysteresis, which is characteristic of a
second-order phase transition. The temperature dependence ($T$) of the
electrical resistivity ($\rho$) exhibits a positive d$\rho$/d$T$ from
300~K to 2~K, indicating the metallic nature of the system over the
entire temperature range studied.

The polar properties of both conventional (insulating) ferroelectrics
and polar metals can be tuned by external factors
such as atomic substitution, chemical ordering and pressure. In a
classical paper, Samara \textit{et al.}~\cite{Samara1975} showed that hydrostatic pressure decreases the transition
temperature in conventional ferroelectrics and ultimately destabilizes
the polarization. That is
because pressure increases short-range interactions much more rapidly
than long-range interactions and as a result, the harmonic
soft-mode frequency becomes less negative with increasing
pressure~\cite{Kornev2007}. Such pressure-induced suppression of ferroelectricity is
observed in BaTiO$_3$~\cite{IshidatePRL1997,ZhuJPCS2008,BousquetPRB2006} and BiFeO$_3$~\cite{HaumontPRB2009}.
On the other hand, negative pressure is found to
increase ferroelectric Curie temperature and polarization in
freestanding PbTiO$_3$ particles~\cite{Wang2015}. 

In this work, we combine transport measurements up to 6.5~GPa and first-principles
calculations to show that hydrostatic (positive) pressure can enhance
the transition temperature of LiOsO$_3$, in contrast to
conventional ferroelectrics. Recently, a novel ``cubic Dirac point" is predicted in the $R\bar{3}c$ phase, which transforms into three mutually crossed nodal rings in the $R3c$ phase \cite{Yu2018}. Thus, the prospect of stabilizing the nodal rings at elevated temperature is attractive.
The pressure-induced enhancement of the transition temperature
is due to the fact that the volume of the
non-centrosymmetric structure ($R3c$) is smaller than that of the
centrosymmetric structure ($R\bar{3}c$). Pressure generically favors
the structure with the smallest volume and therefore with pressure, the polar $R3c$
structure of LiOsO$_3$ becomes more stable over the $R\bar{3}c$
structure and $T_s$ increases. The same analysis
can also be applied to conventional ferroelectrics, which predicts
an increase of $T_C$ and polarization in LiNbO$_3$ 
and ZnSnO$_3$ with pressure~\cite{gu2017arxiv}.

\begin{figure}[!t]\centering
      \resizebox{8.5cm}{!}{
              \includegraphics{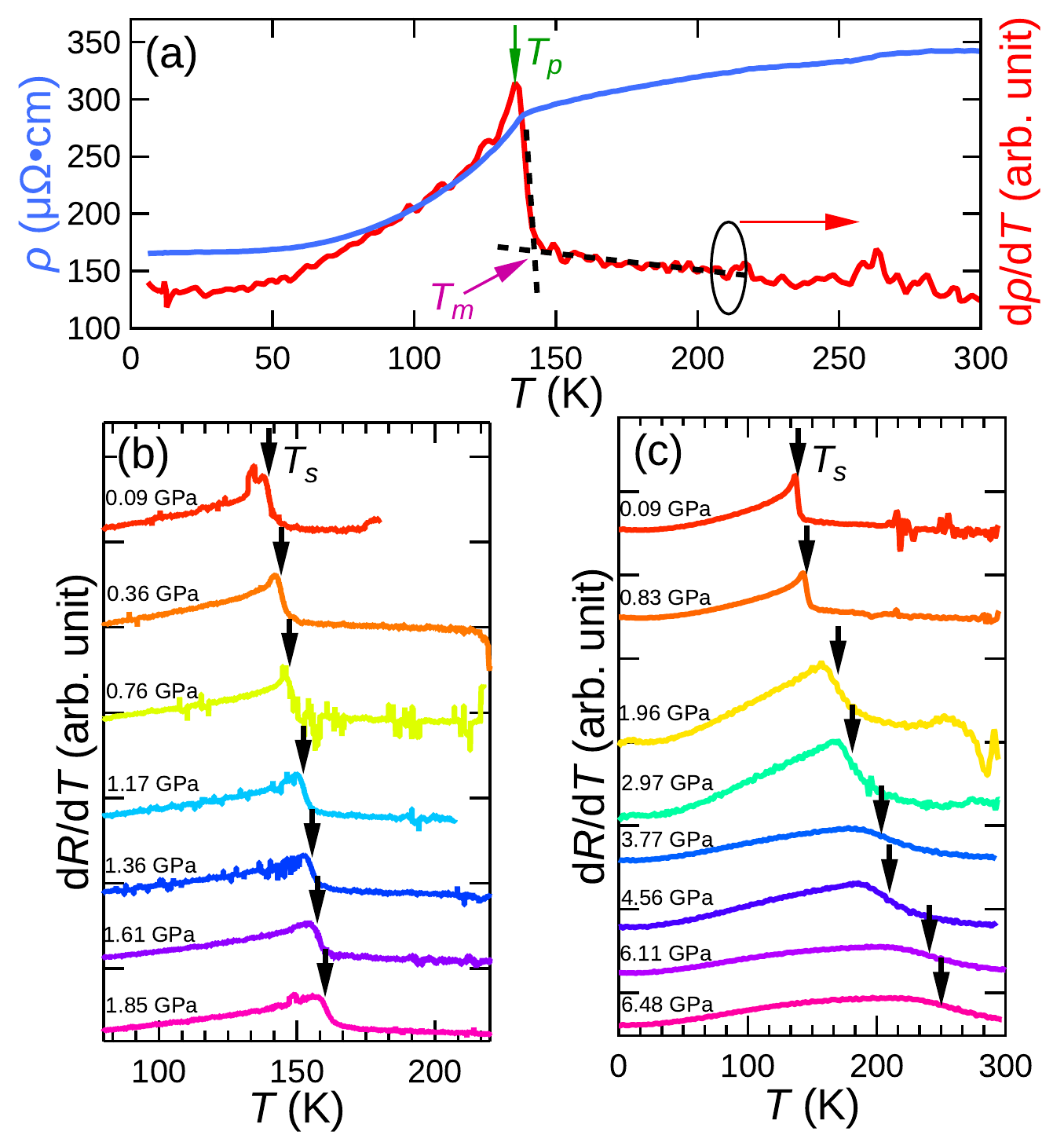}}                				
            \caption{\label{fig1}(a) Temperature dependence of
              $\rho(T)$ and d$\rho$/d$T$ of LiOsO$_3$ at ambient
              pressure. The definitions of $T_p$ and $T_m$ are
              shown. (b) Temperature dependence of the first
              derivative of resistance d$R$/d$T$ of LiOsO$_3$ at
              different pressures provided by the clamp cell. (c)
              Temperature dependence of d$R$/d$T$ of LiOsO$_3$ at
              different pressures in the anvil cell. The vertical
              arrows in (b) and (c) mark $T_s$.}
\end{figure}

Single crystals of LiOsO$_3$ were synthesized by a high-pressure solid
state reaction as described elsewhere \cite{Shi2013}. Electrical
resistance measurements were conducted using a standard four-probe
technique. The electrodes were made with gold wires and silver paste
(Dupont 6838) on the polished surface of the single
crystals. High-pressure experiments were performed in a
piston-cylinder clamp cell and a miniature moissanite anvil cell. The
clamp cell provides a higher pressure-resolution than the anvil cell,
but with a smaller pressure range. Two pieces of single crystals from
the same batch were used, one in each pressure cell. The pressure
achieved in the clamp cell was estimated by the zero-field
superconducting transition of a piece of Pb placed near the sample,
while the pressure value inside the anvil cell was determined by ruby
fluorescence spectroscopy at room temperature. Glycerin was used as
the pressure transmitting medium for both types of pressure
cells. Low-temperature environment down to 2 K was provided by a
Physical Property Measurement System (PPMS) made by Quantum Design.

Density functional theory (DFT) calculations within the
\textit{ab initio} plane-wave approach, as implemented in the Vienna
Ab-initio Simulation Package (VASP)~\cite{Kresse1996,Kresse-PRB-1996},
were performed. A plane wave basis set and projector-augmented wave
method~\cite{Blochl1994} were employed.  PBEsol, a revised
Perdew-Burke-Ernzerhof generalized gradient approximation which
improved equilibrium properties of densely-packed
solids~\cite{Perdew2008}, was used as the exchange-correlation
functional.  The cutoff energy was set as 600~eV and a
$12\times 12\times 12$ Monkhorst-Pack grid was used to
sample the Brillouin zone.  The threshold of energy convergence was
$10^{-7}$ eV.  In the structural relaxation, each
Hellmann-Feynman force component was smaller than $10^{-3}$eV/\AA~ and the stress tensor was smaller than 0.1~GPa. The phonon frequencies at $\Gamma$ point of centrosymmetric ${R\bar{3}c}$ \LiOsO~were computed via density functional perturbation theory method~\cite{Gonze1997}.
 

In LiOsO$_3$, this resistance exhibits a kink at $T_s$. Therefore, the
metallic nature of \LiOsO\ offers an elegant method to follow $T_s$
under pressure using a simple resistance ($R$) measurement, as opposed
to the more elaborate dielectric constant measurement for tracking
$T_C$ in conventional ferroelectric materials.  Figure~\ref{fig1}(a)
shows the temperature dependence of electrical resistivity $\rho(T)$
and its temperature derivative d$\rho$/d$T$ at ambient
pressure. $\rho(T)$ shows a metallic behavior over the entire
temperature range measured, and exhibits a kink at 140~K, consistent
with the previous result \cite{Shi2013}.  d$\rho$/d$T$ shows a rapid
increase near the kink.
We define $T_m$ as the temperature at which d$\rho$/d$T$ begins to
rise rapidly, and $T_p$ as the temperature when d$\rho$/d$T$ peaks (see
Fig.~1(a)). The transition temperature $T_s$ is the average between
$T_m$ and $T_p$. At low pressures, when the kink is pronounced, $T_m$
is very close to $T_p$.  Figure~\ref{fig1}(b) displays the temperature
dependence of the first derivative of resistance d$R$/d$T$ at
different pressures in the clamp cell. With pressure, $T_s$ seemingly
experiences an enhancement. To investigate the behavior of $T_s$ at
higher pressures, the anvil cell data is used, and the temperature
dependence of d$R$/d$T$ is plotted in Fig.~\ref{fig1}(c). At 6.48~GPa,
the highest pressure of this experiment, the peak-like anomaly in
d$R$/d$T$ is significantly broadened. Using $T_p$ and $T_m$, $T_s$ is
determined to be $(250\pm44)$~K at 6.48~GPa. The $R(T)$ curves, from
which the d$R$/d$T$ are constructed, are presented in the
supplementary material.

\begin{figure}[!t]\centering
       \resizebox{8.5cm}{!}{
\includegraphics{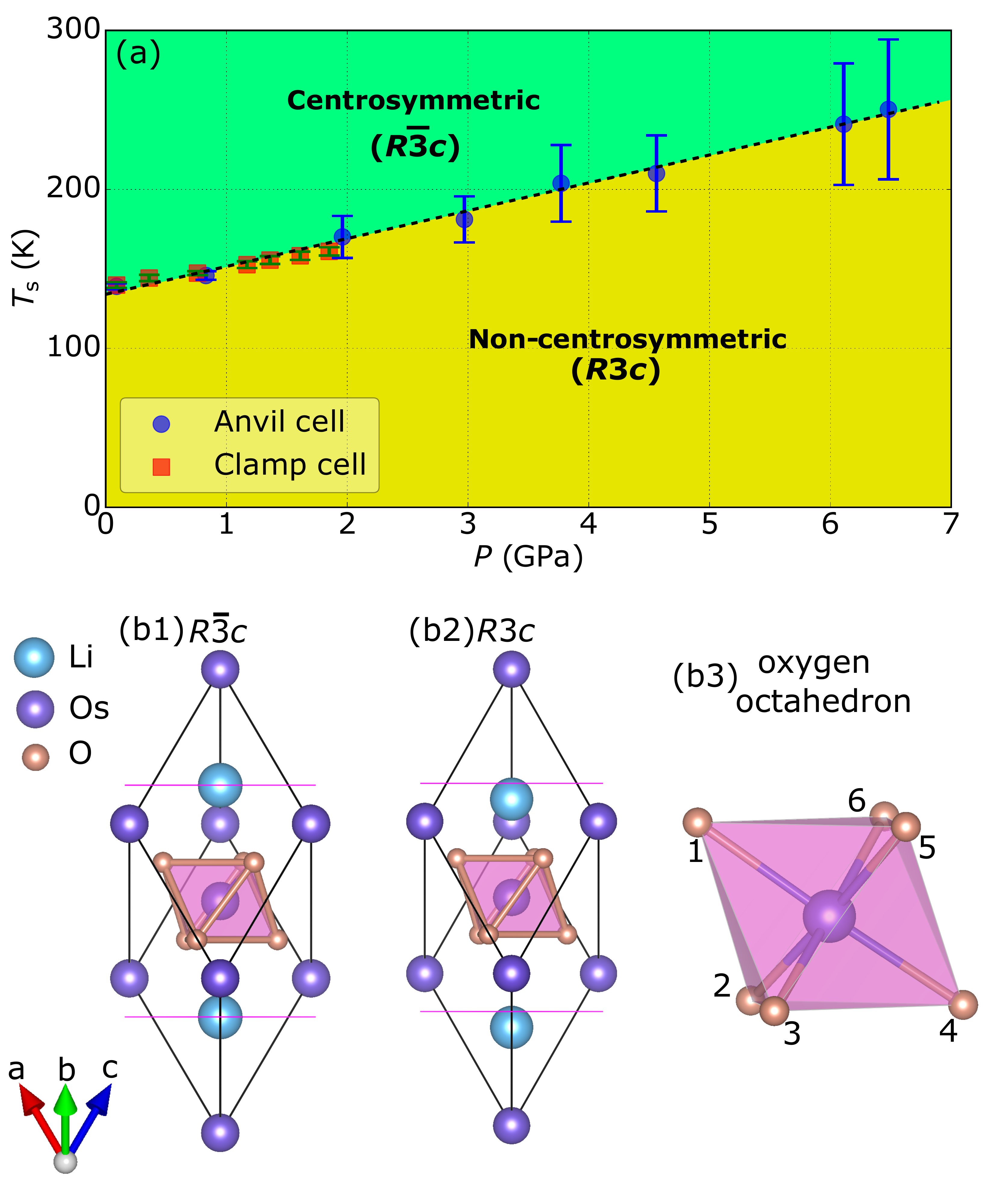}}                				
\caption{\label{fig2} (a) Pressure dependence of $T_s$. The blue dots (red squares)
  with blue (red) error bars represent the data obtained via the anvil (clamp) cell
  technique. The dashed straight line is obtained by a least-squares
  fitting to the anvil cell data. (b) Crystal structures
  of (b1) centrosymmetric ($R\bar{3}c$) and (b2) non-centrosymmetric
  ($R3c$) \LiOsO. The magenta horizontal lines representing the (111)
  planes across the high symmetry points (1/4, 1/4, 1/4) and (3/4, 3/4, 3/4) are shown to guide the eyes. Here, (1/4, 1/4, 1/4) and (3/4, 3/4, 3/4) are fractional coordinates that are parallel to the crystallographic axes. 
  (b3) An OsO$_6$ oxygen octahedron. }
\end{figure}

Figure~\ref{fig2}(a) summarizes the structural transition temperatures $T_s(P)$ at different applied pressures
obtained by the clamp cell and the anvil cell techniques. At the low pressure region ($\leq$2~GPa), $T_s(P)$ obtained via the clamp cell overlaps with that obtained via the anvil cell,
indicating that the data from both techniques are consistent with each other. At the high pressure region ($>$2~GPa),
$T_s(P)$ continues to increase. The overall $T_s(P)$ is linear with a large slope of $\sim$17.54~K/GPa. Therefore,
our experimental results unambiguously demonstrate the enhanced stability of the polar phase.



DFT calculations were performed to understand the observed pressure effect on $T_s$ of LiOsO$_3$.
Figure~\ref{fig2}(b1) and (b2) show the fully
relaxed centrosymmetric structure ($R\bar{3}c$, non-polar)
and the non-centrosymmetric structure ($R3c$, polar) of
\LiOsO, respectively, obtained from DFT-PBEsol calculations at ambient
conditions. In the $R\bar{3}c$ structure, the Os atom is located at
the center of the oxygen octahedron, which is also the symmetric center of
the two Li ions. 
In the ${R3c}$ structure, the inversion symmetry is broken owing to
the dominant contribution from Li displacements and a much smaller contribution
from Os off-center displacements \cite{Giovannetti2014}. 
Figure~\ref{fig2}(b3) displays the OsO$_6$ octahedron in
LiOsO$_3$. In both polar and non-polar structures, the three Os--O
bond angles O1--Os--O4, O2--Os--O5 and O3--Os--O6 are identical,
protected by the $R3c$ and $R\bar{3}c$ symmetries, respectively. Hence
we only focus on the O1--Os--O4 bond angle. This bond angle is 180$^{\circ}$ when protected by $R\bar{3}c$ symmetry but deviates from 180$^{\circ}$ in the $R3c$ symmetry. Therefore, the bond angle is a good structural parameter quantifying the structural change with respect to $R\bar{3}c$ structure. The
experimental~\cite{Shi2013} and theoretical O--Os--O bond angles,
average Os--O bond lengths and volumes of both structures at ambient
conditions are tabulated in Table~\ref{exp-structure}, showing a good agreement between theory and experiment for both $R\bar{3}c$ and $R3c$ symmetries.
More importantly, the volume of polar structure ($R3c$) is smaller than that of non-polar structure  ($R\bar{3}c$). This will play an essential role in understanding the enhancement of $T_s$ under pressure.

\begin{table}[!t]
\centering
\caption{Structural parameters for $R\bar{3}c$ and ${R3c}$ structures of LiOsO$_3$ obtained from
  DFT calculations, compared to experimental values~\cite{Shi2013}
	shown in brackets. The predicted ${Pbnm}$ structure which is next to ${R\bar{3}c}$ LiOsO$_3$ in energy at 0~GPa is also shown as a comparison, more details on ${Pbnm}$ phase can be found in the supplementary material. 
  The volume is normalized to per formula unit of LiOsO$_3$, the bond angle corresponds to
  angle O1--Os--O4 in Fig.~\ref{fig2}(b3), and the bond length corresponds to the average length of
  of O--Os in Fig. \ref{fig2}(b3).}
\label{exp-structure}
\begin{tabular}{c | cccc}
	\hline\hline
	structural & \multicolumn{4}{c}{space group} \\
	parameter & & ${R\bar{3}c}$  &   ${R3c}$ & ${Pbnm}$\\
	\hline
		volume (\AA$^3$) & &  49.12 (48.90) & 48.71 (48.65) & 48.45 \\
		bond angle ($^{\circ}$)  & & 180 (180) & 178.03 (176.84) & 180\\
		bond length (\AA) & & 1.946 (1.944) & 1.947 (1.944) & 1.946\\
                \hline\hline
\end{tabular}
\end{table}

\begin{figure}[!b]\centering
       \resizebox{8.0cm}{!}{
              \includegraphics{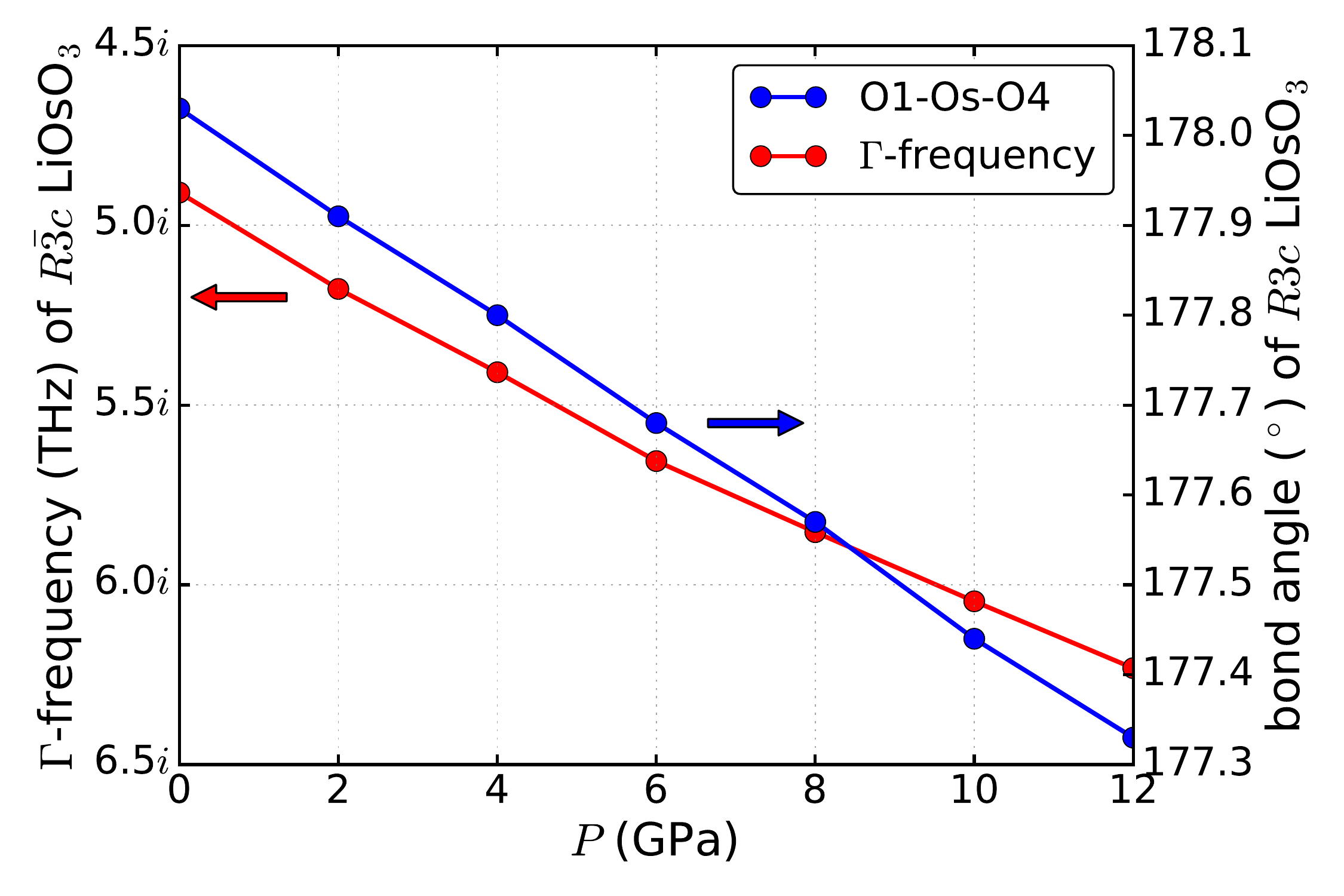}}
              \caption{\label{fig3} (Left vertical axis) Lowest
                zone-center phonon frequency of $R\bar{3}c$ LiOsO$_3$ as a function of pressure. (Right vertical axis) O1--Os--O4 bond
              angle of $R3c$ LiOsO$_3$ as a function of pressure. The O1--Os--O4 bond angle
              is shown in Fig.~\ref{fig2}(b3).}
\end{figure}


The left axis in Fig. \ref{fig3}~shows the pressure dependence of the lowest phonon frequency
at $\Gamma$ point of the non-polar ${R\bar{3}c}$
LiOsO$_3$ at zero temperature. The phonon frequency is found to be imaginary and its magnitude increases with pressure, indicating 
that the non-polar $R\bar{3}c$ structure is unstable and pressure enhances the instability of the 
zone-center phonon mode. On the other hand,
the right axis in Fig.~\ref{fig3} shows the pressure dependence of the O1--Os--O4 bond angle of polar $R3c$ LiOsO$_3$. 
As pressure increases, the bond angle decreases and deviates further 
from the ideal 180$^{\circ}$. This shows that applying pressure can enhance the polar nature of ${R3c}$ LiOsO$_3$.
Therefore, both results show that external pressure not only destabilizes non-polar ${R\bar{3}c}$ LiOsO$_3$, but also
renders ${R3c}$ LiOsO$_3$ more polar.


\begin{figure}[!t]\centering
       \resizebox{8cm}{!}{
              \includegraphics{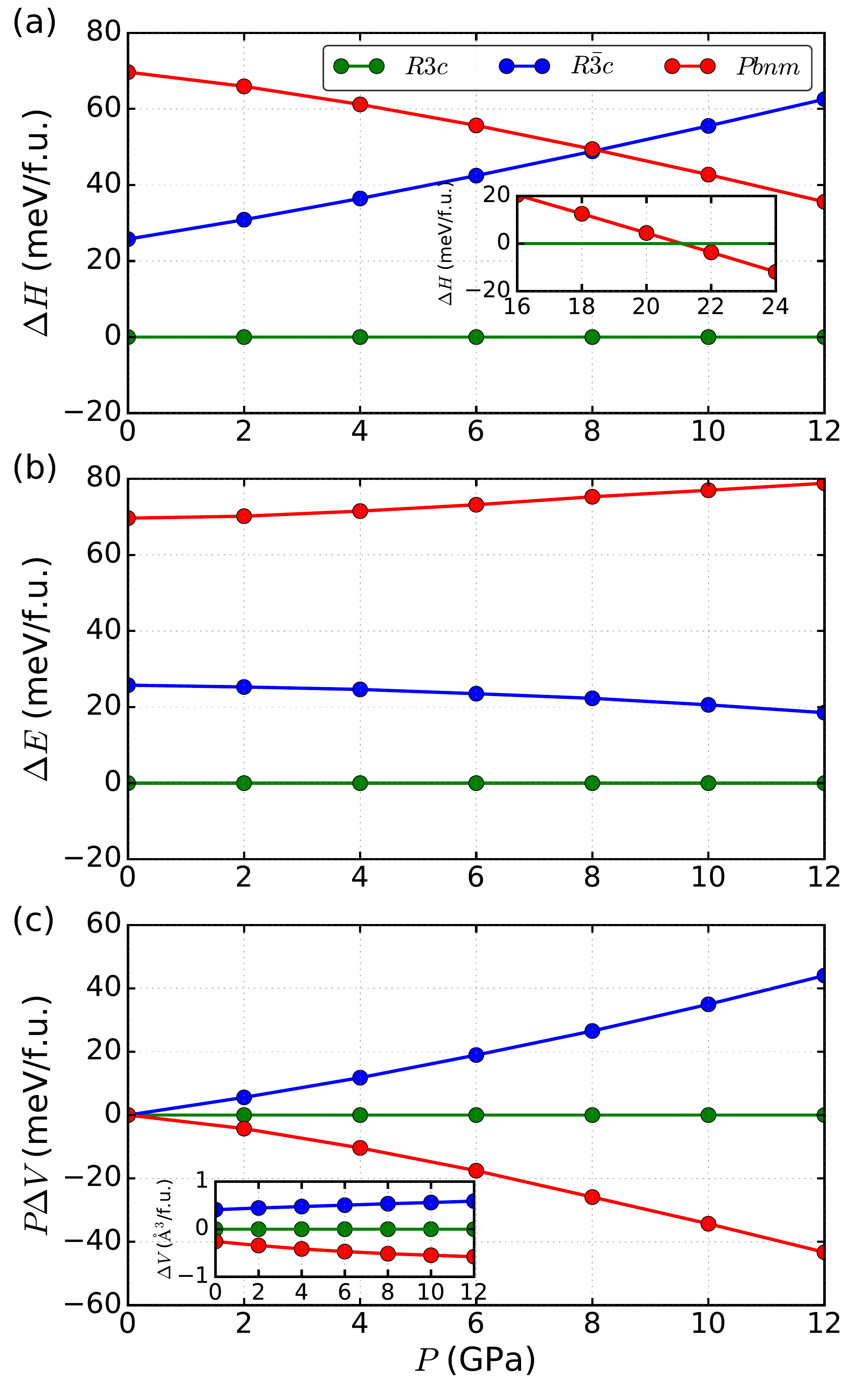}}
              \caption{\label{fig4} Zero-temperature thermodynamic
properties of LiOsO$_3$ under pressure. (a) Enthalpy difference $\Delta H$ between ${R3c}$ and other structures. The inset in (a) shows the $\Delta H$ under 16 $\sim$ 24~GPa.
(b) Energy difference $\Delta E$ between ${R3c}$ and other structures.
(c) Pressure times
volume difference $P\Delta V$ between $R3c$ and other structures. 
The inset in (c) shows $\Delta V$ between $R3c$
and other structures. The green, blue and red
symbols correspond to ${R3c}$, $R\bar{3}c$ and ${Pbnm}$
structures, respectively. ${R3c}$ is chosen as the reference state.}
\end{figure}

To understand why pressure favors the polar structure $R3c$ over the
non-polar structure $R\bar{3}c$, zero-temperature thermodynamic
functions of LiOsO$_3$ were computed via DFT calculations.
Figure~\ref{fig4}(a) shows the zero-temperature enthalpy $H$ of
LiOsO$_3$ in three structures with the lowest enthalpy, $R3c$,
$R\bar{3}c$ and $Pbnm$, based on our structure search study (see the
supplementary material).  For all pressures, the enthalpy of the
$R3c$ structure is used as the reference. The blue and green symbols
in Fig.~\ref{fig4}(a) show that the $R3c$ structure is more stable
than the $R\bar{3}c$ structure, and pressure increases the stability
of the $R3c$ structure over the $R\bar{3}c$ structure, which is
consistent with the trend shown in Fig.~\ref{fig3}.  On the other
hand, though the $Pbnm$ structure has higher energy than the $R3c$ structure
at ambient pressure, applying pressure reduces the enthalpy difference
between $Pbnm$ and $R3c$ structures, and eventually stabilizes the
$Pbnm$ structure over the $R3c$ structure at about 21~GPa.  It is
known that the ferroelectric transition temperature is proportional to
the energy difference between the polar structure and the
corresponding high-symmetry non-polar
structure~\cite{Wojde2013arxiv}. Under pressure, enthalpy replaces
energy as the thermodynamic function. Therefore, the increasing
enthalpy difference between $R3c$ and $R\bar{3}c$ structures leads to
the enhancement of $T_s$, which is consistent with our
experiment. Furthermore, the enthalpy difference between $R3c$ and
$R\bar{3}c$ structures, and the Li displacement squared in ${R3c}$ 
LiOsO$_3$ (see the supplementary material) almost increase linearly with pressure, which
is also consistent with the experimentally observed linear $T_s(P)$
displayed in Fig.~\ref{fig2}(a).

Using the relation $H=E+PV$, the total energy $E$ and $PV$ for
different structures of LiOsO$_3$, with respect to $R3c$, are
illustrated in Figs.~\ref{fig4}(b) and (c), respectively. As pressure
increases, $\Delta E$ only changes slightly. Moreover, the pressure
dependence of $\Delta E$ is opposite to that of $\Delta H$.  This
indicates that the pressure dependence of $\Delta H$ is dominated by
$P\Delta V$. As the inset of Fig.~\ref{fig4}(c) shows, the volume of
the polar $R3c$ LiOsO$_3$ is always smaller than that of the non-polar
$R\bar{3}c$ LiOsO$_3$, and $\Delta V$ is almost a constant within the
pressure range we study. As a result, $P \Delta V$ increases linearly
with $P$ and so does $\Delta H$ because $\Delta E$ has weak dependence
on $P$.

Figure~\ref{fig4} provides a simple criterion to predict the pressure 
effects on polar metals. If the volume of the polar structure is 
smaller than that of the non-polar structure at ambient pressure,
pressure enhances the polar properties 
(increasing $T_s$ and polar distortions). Conversely, if the volume of polar structure is larger, pressure 
will suppress the polar properties (decreasing $T_s$). 
This simple criterion can also be applied to conventional insulating 
ferroelectrics~(see supplementary material). For example, LiNbO$_3$ and ZnSnO$_3$ belong 
to the former class ($V(\textrm{polar}) < V(\textrm{non-polar})$), which have been predicted to feature a pressure-enhanced ground state polarization and $T_C$~\cite{xiang2017prb,gu2017arxiv}. On the other hand, BaTiO$_3$ and BiFeO$_3$ belong to 
the latter class ($V(\textrm{polar}) > V(\textrm{non-polar})$), and it has been observed that  
pressure reduces the polarization and eventually destabilizes the
ferroelectric ground state~\cite{IshidatePRL1997,ZhuJPCS2008,BousquetPRB2006,HaumontPRB2009}.


Note that the enhancement of $T_s$ via pressure 
in LiOsO$_3$ will be thwarted by the stabilization of a different
structure. As Fig.~\ref{fig4}(a) shows, the non-polar $Pbnm$ structure 
becomes more stable than the non-polar $R\bar{3}c$ structure above 8~GPa, 
and becomes more stable than the polar $R3c$ structure above 21~GPa.
As a conservative estimation using the linear pressure dependence,
$T_s$ in LiOsO$_3$ can be enhanced up to $\sim$274 K at 8~GPa. The robustness of our conclusions is also tested by considering the
effects of Hubbard $U$, spin-orbital coupling (SOC) and possible 
magnetic ordering. The results from DFT+$U$, DFT+SOC and 
DFT+$U$+SOC calculations all agree qualitatively with 
Fig.~\ref{fig4}(a) (see the supplementary material for details).

In summary, the pressure dependence of structural transition
temperature $T_s$ in polar metal LiOsO$_3$ was measured up to
$\sim6.5$~GPa. $T_s$ increases linearly under pressure, reaching 250~K
at $\sim6.5$~GPa.  The experimental result is corroborated by
first-principle calculations which show that pressure further 
stabilizes the polar $R3c$ phase over the centrosymmetric $R\bar{3}c$ 
phase and enhances the polar properties in the $R3c$ structure.
The enhancement of $T_s$
arises from the fact that the volume of the polar $R3c$ structure is
smaller than that of the non-polar $R\bar{3}c$ structure, and pressure
generically favors the structure with the smallest volume. This
criterion can also be applied to insulating
ferroelectrics, which predicts pressure increases polarization and 
$T_C$ in ZnSnO$_3$ and LiNbO$_3$~\cite{gu2017arxiv}.
\\

See supplementary material for the method for determining $T_s$ and further details on theoretical investigations including the structure search and the connection between the volume and oxygen octahedra rotations. 
\\

We acknowledge support from
Research Grants Council of Hong Kong (ECS/24300214, GRF/14300117),
CUHK Direct Grant (4053223, 4053299), National Natural Science
Foundation of China (11774236, 11504310), the Seed Grants of
NYU-ECNU Joint Research Institutes, the JSPS KAKENHI (JP15K14133 and
JP16H04501), and JSPS Bilateral Open Partnership Joint Research
Projects. We thank Sheng Wang for helpful discussions
on crystal structure prediction. Extreme Science and
Engineering Discovery Environment (XSEDE) provides computational resources.

%




%
%

\end{document}